\documentclass[%
 preprint,
reprint,
superscriptaddress,
 amsmath,amssymb,
 aps,
floatfix,
]{revtex4-2}

\usepackage{graphicx}
\usepackage{float}
\usepackage{subfig}
\usepackage{dcolumn}
\usepackage{bm}
\usepackage{xcolor}
\usepackage{soul}

\newcommand{\fixes}[2]{#1}

\begin{document}

\preprint{APS/123-QED}

\title{Accelerating Training of MLIPs Through Small-Cell Training}

\author{Jason A. Meziere}
\email{jmezier2@byu.edu}
\affiliation{Department of Physics, Brigham Young University, Provo UT 84602 USA}
\author{Yu Luo}
\email{yu.luo@queensu.ca}
\affiliation{Department of Mechanical \& Materials Engineering, Queen’s University, Kingston, Ontario, K7L 2N8, Canada}
\author{Yi Xia}
\email{yxia@pdx.edu}
\affiliation{Department of Mechanical \& Materials Engineering, Portland State University, Portland, OR 97201, USA}
\author{Laurent Karim Béland}
\email{laurent.beland@queensu.ca}
\affiliation{Department of Mechanical \& Materials Engineering, Queen’s University, Kingston, Ontario, K7L 2N8, Canada}
\author{Mark R. Daymond}
\email{mark.daymond@queensu.ca}
\affiliation{Department of Mechanical \& Materials Engineering, Queen’s University, Kingston, Ontario, K7L 2N8, Canada}
\author{Gus L. W. Hart}%
\email{gus.hart@byu.edu}
\affiliation{Department of Physics, Brigham Young University, Provo UT 84602 USA}%

\date{\today}

\begin{abstract}

While machine-learned interatomic potentials have become a mainstay for modeling materials, designing training sets that lead to robust potentials is challenging. Automated methods, such as active learning and on-the-fly learning, \fixes{construct}{allow for the construction of} reliable training sets, but these processes can be \fixes{}{very }resource-intensive\fixes{}{ when training a potential for use in large-scale simulations}. Current training approaches often use density functional theory (DFT) calculations that have the same cell size as the simulations that \fixes{the potential is explicitly trained to model}{use the potential}. Here, we demonstrate an easy-to-implement small-cell \fixes{}{structures }training protocol and use it to \fixes{model the Zr-H system}{train a potential for zirconium and hydrides}. This training leads to a \fixes{potential that accurately predicts known stable Zr-H phases and reproduces the $\alpha$-$\beta$ pure zirconium phase transition in molecular dynamics simulations}{convex hull in good agreement with DFT when applied to known stable phases}. Compared to traditional active learning, small-cell training decreased the training time of \fixes{}{a potential able to capture }the $\alpha$-$\beta$ \fixes{zirconium }{}phase transition by approximately 20 times. The potential describes the phase transition with a degree of accuracy similar to that of the large-cell training method. 

\end{abstract}

\maketitle


\section{Introduction}

In the past decade, machine learned interatomic potentials (MLIPs) have become an indispensable tool for modeling materials because MLIPs expand the length and time scales that can be modeled. These potentials fill the gap between density functional theory (DFT) and fast, physically motivated models such as embedded atom model (EAM) interatomic potentials \cite{deringer_machine_2019, lysogorskiy_performant_2021}. 

MLIPs are created by fitting flexible, parameter-rich models to DFT data \cite{shapeev_moment_2016, behler_neural_2011, szlachta_accuracy_2014, bartok_gaussian_2010}. These potentials can reproduce energies, forces, and stresses for structures nearly as accurately as DFT, but are orders of magnitude faster \cite{vita_exploring_2021}. MLIPs simulate longer time scales than possible with DFT, successfully modeling material properties such as crack propagation and phase transitions or generating phase diagrams \cite{bartok_representing_2013, rosenbrock_machine-learned_2021}. On the other hand, EAMs have a simpler, physically motivated, mathematical form \cite{daw_semiempirical_1983, daw_embedded-atom_1984}, which makes them faster and more robust than MLIPs, but far less flexible. EAMs have tens of parameters while MLIPs can have thousands or tens of thousands. EAMs do not reproduce diverse chemical behavior because of their small number of parameters \cite{takahashi_conceptual_2017}. While EAM potentials have been able to model complex phenomena (e.g. in the Zr-H system \cite{wimmer_hydrogen_2020, starikov_optimized_2021, nicholls_transferability_2023}), MLIPs can more accurately model material properties such as phonon bandstructures, elastic constants, point defects and planar defects \cite{bartok_machine_2018, takahashi_conceptual_2017}.

Despite MLIPs' successes, creating accurate and robust potentials is still challenging \cite{podryabinkin_active_2017}. A primary difficulty in creating an effective MLIP is determining a sufficient, but efficient, training set \cite{hodapp_machine-learning_2021, bartok_representing_2013}. In order for the MLIP to accurately predict a structure's energetics, the training set must contain atomic environments representative of that structure, but knowing which structures the MLIP will need to model before the training process begins is impossible. \textit{A priori} construction of a training set relies on intuition rather than mathematical formalism and cannot guarantee that complex material properties will be modeled accurately. 

\begin{figure*}
  \centering
  \captionsetup{justification=raggedright}
  \subfloat[$\alpha$ phase]{\raisebox{-0.5\height}{\includegraphics[width=0.25\linewidth]{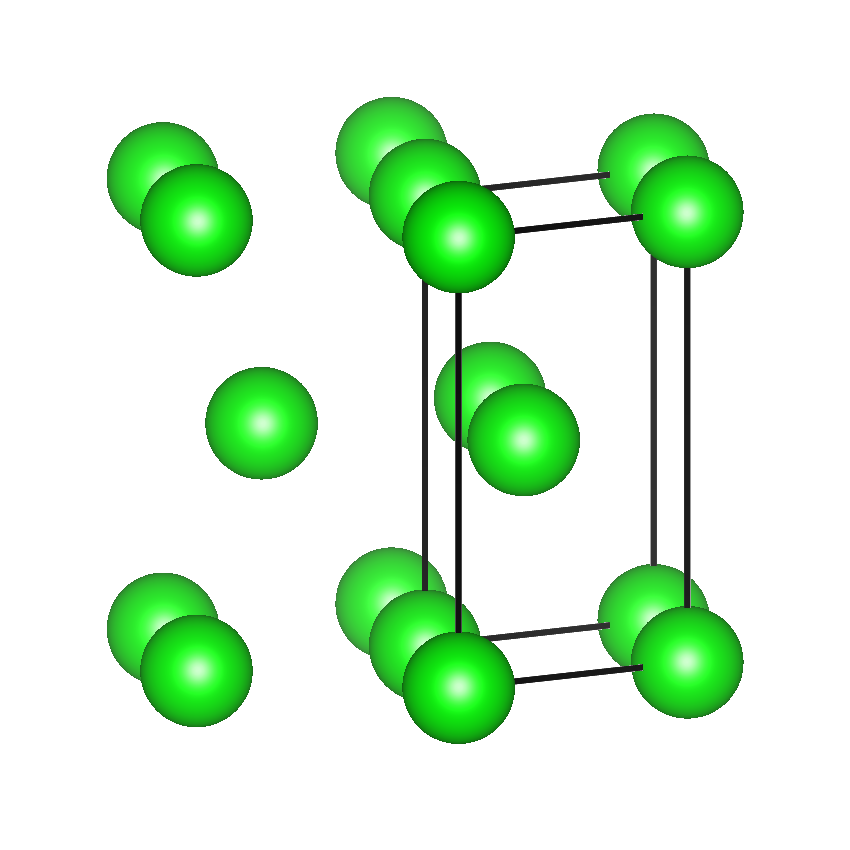}}}
  \hfill
  \subfloat[$\delta$ phase]{\raisebox{-0.5\height}{\includegraphics[width=0.25\linewidth]{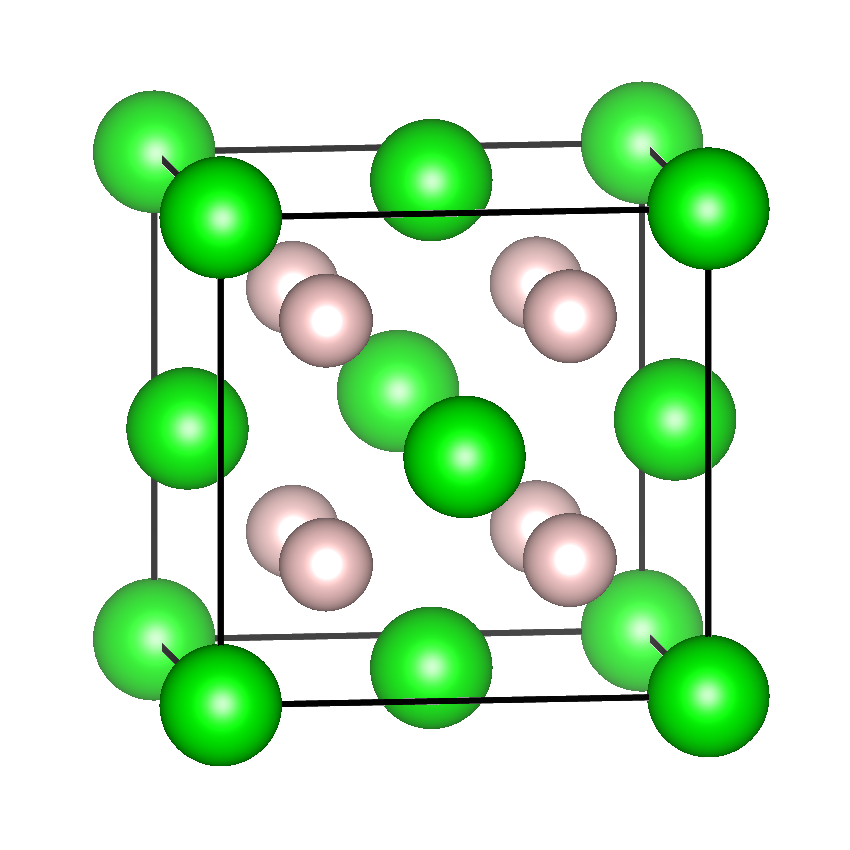}}}
  \hfill
  \subfloat[$\epsilon$ phase]{\raisebox{-0.5\height}{\includegraphics[width=0.25\linewidth]{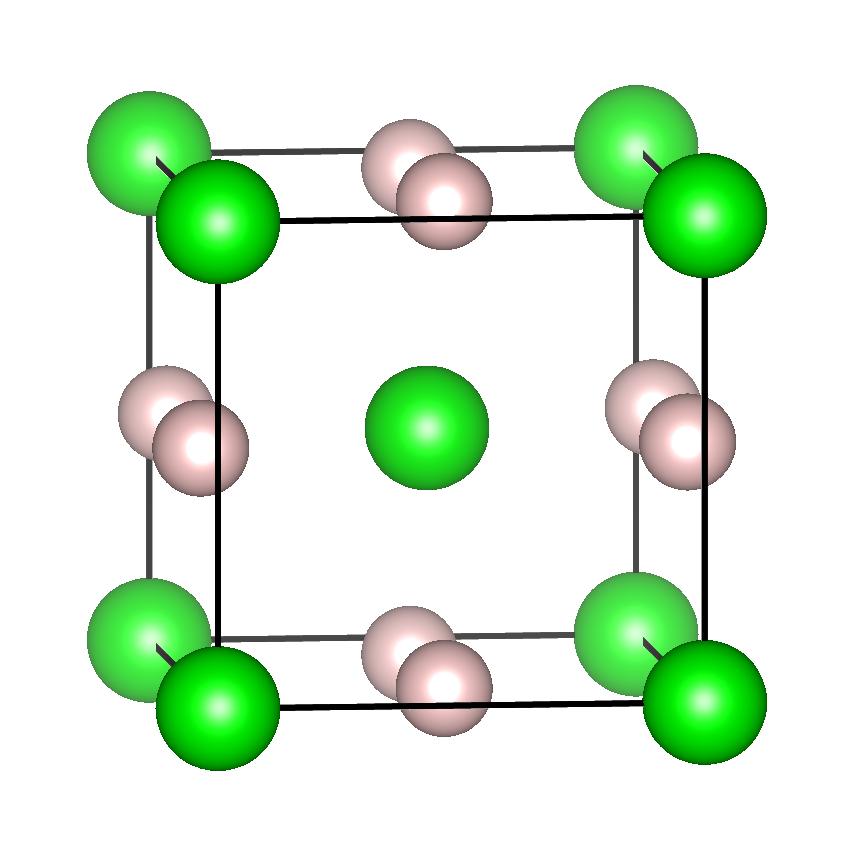}}}
  \hfill
  \subfloat[$\gamma$ phase]{\raisebox{-0.5\height}{\includegraphics[width=0.25\linewidth]{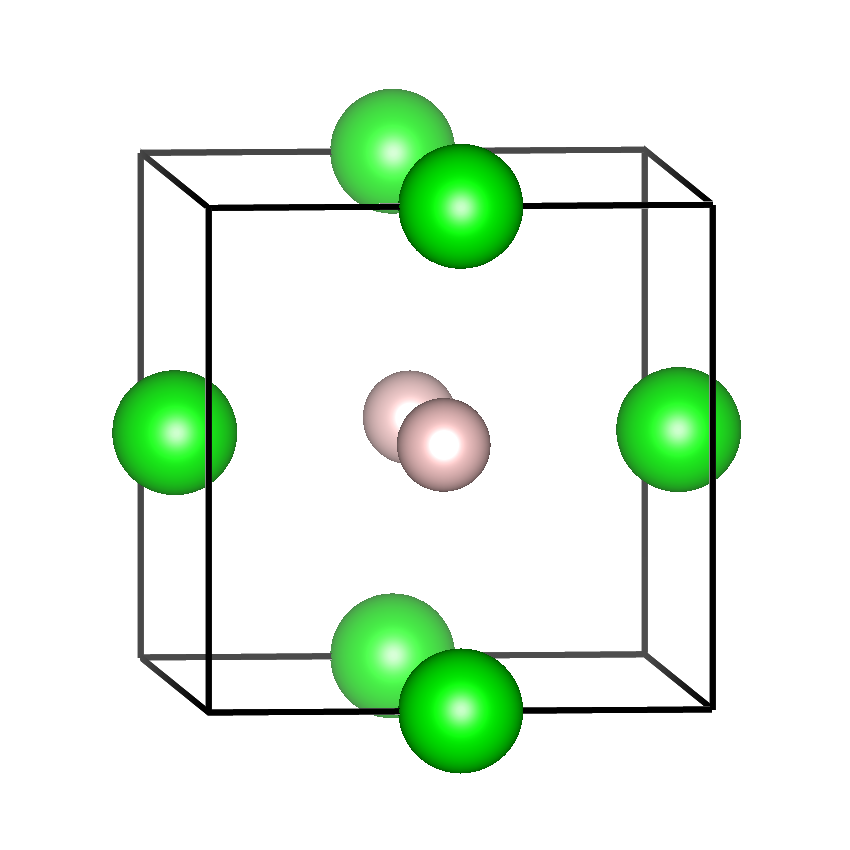}}}
\caption{\label{fig:structs}The $\alpha$, $\delta$, $\epsilon$, and $\gamma$ phases are common pure Zr or Zr-H systems that have been experimentally measured. Above, green atoms are zirconium atoms and red atoms are hydrogen atoms.}
\end{figure*}

There are two common approaches that leverage iterative, rather than \textit{a priori} construction, of the training set: active learning and on-the-fly learning \cite{podryabinkin_accelerating_2019, zhang_active_2019, novikov_mlip_2021}. Both methods use a metric that indicates the level of confidence in the MLIP’s predictions. If the level of confidence in a prediction is insufficient, energies and forces of that structure are calculated with DFT and added to the training set. In this way, active learning and on-the-fly learning iteratively improve the training set until the resulting potential reproduces the material properties of interest \cite{shapeev_active_2020, podryabinkin_active_2017}. 

The philosophy of active learning differs from on-the-fly learning. On-the-fly learning continues trying to improve the potential whenever in use, so the potential is never ``finished''. Active learning, on the other hand, uses a predefined set of simulations to conduct training. Once the potential successfully finishes all simulations in the predefined set without triggering more DFT calculations, the potential is considered reliable and training ends. 

While active learning robustly provides training sets that sufficiently represent desired material properties, typical training approaches are inefficient, using large-cell structures to train potentials that \fixes{model large-cell material properties (material properties that only manifest in simulations using large unit cells containing many atoms)}{will be used in large-cell simulations}. \fixes{We demonstrate that a protocol which prioritizes training on small-cell structures allows information gained from the small-cell structure training to be used when modeling the larger-cell structures.}{We demonstrate below that training with sufficiently diverse \emph{small-cell} structures can yield potentials that are reliable in large scale simulations.} Other recent work has also suggested the small-cell training approach\cite{pickard_ephemeral_2022}, and below we outline a systematic approach. \fixes{}{They typically do not use small-cell structures to capture large-cell structural information. We demonstrate that a protocol which prioritizes training on small-cell structures allows information gained from the small-cell structure training to be used when modeling the larger-cell structures.}

We demonstrate small-cell training with the moment tensor potential (MTP) as implemented in the MLIP-2 package, which includes active learning \cite{shapeev_moment_2016}. MTPs use an invariant polynomial basis to represent atomic densities. MLIP-2 employs the maximal volume (MV) grade \cite{shapeev_active_2020, podryabinkin_active_2017} to determine if a configuration should be added to the training set or not. While we use the MTP/MLIP-2 framework, the methods we describe can be applied to other MLIPs that utilize active learning.

\begin{table}
\scriptsize{}
\captionsetup{justification=raggedright}
\centering
\begin{tabular}{c | c | c | c | c | c | c }
 & a (\AA{}) & b (\AA{})  & c (\AA{})  & $\alpha$ (deg.) & $\beta$ (deg.) & $\gamma$ (deg.) \\\hline
 $\alpha$ (VASP) & 3.230 & 3.230 & 5.167 & 90.0 & 90.0 & 120.0 \\
 $\alpha$ (MTP) & 3.228 & 3.228 & 5.171 & 90.0 & 90.0 & 120.0 \\\hline
 $\delta$ (VASP) & 3.395 & 3.395 & 3.395 & 60.0 & 60.0 & 60.0 \\
 $\delta$ (MTP) & 3.391 & 3.391 & 3.391 & 60.0 & 60.0 & 60.0 \\\hline
 $\epsilon$ (VASP) & 3.331 & 3.331 & 3.331 & 116.0 & 116.0 & 97.1 \\
 $\epsilon$ (MTP) & 3.332 & 3.332 & 3.332 & 115.9 & 115.9 & 97.2 \\\hline
 $\gamma$ (VASP) & 3.236 & 3.236 & 4.996 & 90.0 & 90.0 & 90.0 \\
 $\gamma$ (MTP) & 3.226 & 3.226 & 5.020 & 90.0 & 90.0 & 90.0 \\\hline
\end{tabular}
\caption{\label{tab:relax_prop_initial}Relaxed structure properties for \fixes{the primitive}{} $\alpha$, $\delta$, $\epsilon$, and $\gamma$ phases. Graphically, these phases are shown in Fig.~\ref{fig:structs}}
\end{table}

\section{Z\lowercase{r}-H System}

Many of zirconium's industrial uses are related to nuclear power applications because of its low neutron cross-section. Most current commercial reactor designs include fuel cladding made of zirconium alloys. Furthermore, Canadian Deuterium Uranium (CANDU) reactors employ zirconium alloys as a structural material; CANDU pressure tubes and calandria tubes are made of zirconium alloys. There are two allotropic forms of pure zirconium, $\alpha$-Zr (HCP) and $\beta$-Zr (BCC), which is the high temperature phase.  Water chemistry on the primary side (reactor core) and indeed on the secondary side (steam generation) of nuclear power plants is typically controlled to provide highly reducing chemical conditions to favor the formation of protective metal oxide layers. Therefore, oxidation reaction in metals (anode) exposed to the water will be electrochemically balanced (cathode) by reduction of water and generation of hydrogen. Some fraction of this hydrogen is picked up by the zirconium alloys \cite{kass1960hydrogen}. At typical reactor operating temperatures ($\sim$573K) and hydrogen concentrations, the hydrogen will be in solid solution with the zirconium alloy. However, as reactors are powered down, the temperature decreases, which reduces the solubility limit of hydrogen. This may lead to precipitation of zirconium hydrides, causing embrittlement and possible failure of the material, which is particularly a concern during storage of spent-fuel bundles \cite{azevedo_selection_2011}. Further, hydrogen diffuses `up' hydrostatic tensile stress gradients \cite{Kammenzind2000} leading to elevated concentrations at notches or defects, and hence to the slow-cracking phenomenon of delayed hydride cracking \cite{kearns1967terminal}, an issue of concern for nuclear power plant operators. Therefore, understanding how hydrogen pickup and hydride formation occur in zirconium are considered important for nuclear reactor safety \cite{duan_current_2017,  motta_hydrogen_2019}.

\begin{table}
\scriptsize{}
\centering
\sffamily
\begin{tabular}{ l | l }
 Parameter & Value \\ \hline
 PREC & Accurate \\
 ENCUT & 400 \\
 NELMIN & 2 \\
 NELM & 30 \\
 ISMEAR & 0 \\
 SIGMA & 0.05 \\
 ISIF & 3 \\
 EDIFF & 1.0E-06 \\
 EDIFFG & -0.01 \\
 VOSKOWN & 1 \\
 NBLOCK & 1 \\
 NWRITE & 1 \\
 NELM & 60 \\
 ALGO & Normal \\
 ISPIN & 1 \\
 INIWAV & 1 \\
 ISTART & 0 \\
 ICHARG & 2 \\
 LWAVE & .FALSE. \\
 LCHARG & .FALSE. \\
 ADDGRID & .FALSE. \\
 LREAL & .FALSE. \\
 NCORE & 1 \\
\end{tabular}
\caption{\label{tab:vasp}Parameters used in all VASP calculations.}
\end{table}

The crystal structure of the hydrides precipitated in zirconium is dependent on hydrogen concentration. At the relatively low hydrogen concentrations found in operational reactors, face-centered cubic $\delta$-hydride\fixes{}{s} is the most stable and most frequently observed hydride phase in most commercial zirconium alloys \cite{ells1968hydride}. Meanwhile, face-centered tetragonal $\gamma$-hydrides are sometimes observed experimentally in commercial alloys \cite{daum2009identification} and are commonly seen in pure zirconium \cite{Long2021}. At higher hydrogen concentrations, face-centered tetragonal $\epsilon$-hydrides are also reported \cite{beck1962zirconium}. These phases are shown in Fig.~\ref{fig:structs}.

The specific case of phase transformation of zirconium hydrides is of particular interest. Despite decades of research, there is no scientific consensus as to the path of nucleation and growth mechanisms and the role that different hydride phases may play in the underlying precipitation of hydrides, with discussions dating to the 1970s \cite{Carpenter1978} and continuing today \cite{McRae2018Precipitates, Badr2023Trigonal}.  Further, experimental evidence strongly suggests that the pathway (whether thermodynamically or kinetically driven) for formation of $\gamma$ versus $\delta$ hydride phases is finely balanced, with small changes sufficient to favor one over the other \cite{Long2021,lanzani_comments_2004,Tulk2012}. Yet, the relative likelihood as to which phase forms has practical application in predicting performance in-reactor due to the phases' quite different volume expansions compared to zirconium \cite{Carpenter1973} and hence impact on component behavior.  From an atomistic simulation perspective, no current classical interatomic interaction potential is able to describe the structure and energetics of the multiple Zr-H phases, which limits atom-scale studies of this system to the small length and time scales accessible to DFT, such as in Refs. \cite{njifon2021first,domain2002atomic,udagawa2010ab,zhu2010first,wang2012first,olsson2014ab,zhu2018structure,christensen2015h,huang2019first}. The development of a classical potential able to capture the main Zr-H phases is an important step towards a comprehensive understanding of the formation thermodynamics and kinetics of zirconium hydrides.

\section{Small-cell Structure Training Protocol}

All active learning methods are iterative in nature. Training structures are systematically added, based on actual or predicted error, to improve the accuracy of potential. During the training process, the potential is tested for a variety of situations, errors assessed, and training structures that contribute the most new information are calculated with DFT and added to the training set. This cycle continues until the potential passes all required robustness tests. In situations where the potential will be used for large-scale simulations (say more than 100 atoms), such as modeling structural phase transitions, large-cell structures are typically included in the training process. In contrast, we demonstrate that a reliable potential is possible with training sets that contain only small-cell structures.

In the small-cell approach, the first training set contains primitive unit cells of experimentally relevant phases. It is usually helpful to include a few other configurations to provide basic physics for the training. For example, primitive cells at several volume\fixes{s}{} provide basic equation of state information.

In the second step, slightly larger unit cells, modifications of the experimentally relevant phases with varying concentrations, are tested with the initial potential. These can be systematically enumerated by cell size \cite{hart_algorithm_2008, hart_generating_2009}, providing a methodical approach to choosing the next set of structures to test the potential with. Structures that are not accurately handled by the potential are calculated in DFT, added to the training set, and the potential refitted. The potential is then tested on the enumerated structures of the next-largest cell-size. Training stops when the potential effectively handles all the new structures in the added set.

In the case of MTP, a convenient test for the potential is to minimize the energy of a test structure with respect to its unit cell shape, size, and atomic position\fixes{s}{}. Training stops when the predicted error is sufficiently low for the entire structural ``relaxation trajectory'' for all of the enumerated structures in a test set. How to test (and improve) this potential (if necessary) for large-scale simulations will be demonstrated below.

We initially developed a Zr-H potential to search for thermodynamically stable structures over a large concentration range of hydrogen. Knowing the stable structures of the Zr-H system is a first step for understanding the energetic landscape of possible structures that occur in a nuclear environment. Currently, the structures considered to be thermodynamically stable at room temperature in the Zr-H system are the pure zirconium $\alpha$ phase, the $\gamma$ (Zr-H) phase, the the $\delta$ (Zr-H\textsubscript{1.6-1.7}) phase and the ZrH\textsubscript{2} $\epsilon$ phase. There is also a metastable $\beta$ (pure zirconium) phase \cite{lanzani_comments_2004, zhang_understanding_2019}. While the ambient metastability of $\beta$-Zr is not in question, there continues to be extensive discussion in the literature as to the relative stability of the $\delta$ and $\gamma$ phases when the H to Zr ratio is smaller than 1.0, with phase formation experimentally dependent on thermal history as well as the properties (e.g., dislocation density and minority chemistry) of the zirconium, as reviewed in \cite{Long2021}; we note in particular that for hydrides formed from high purity Zr when the H to Zr ratio is smaller than 1.0, $\gamma$ is stable at room temperature, while from even somewhat impure Zr it is $\delta$ which appears stable \cite{Long2021}.  The most widely accepted phase diagram for the pure ZrH binary \cite{Zuzek2000} reports $\gamma$ as a room temperature stable phase  when the H to Zr ratio is smaller than 1.0 and the temperature is below 286$^{\circ}$C. $\delta$ is considered stable at higher H to Zr ratios and at higher temperatures; this is in contrast to earlier phase diagrams \cite{Zuzek1990}.

\begin{figure}
  \centering
  \captionsetup{justification=raggedright}
  \subfloat[]{\includegraphics[width=0.45\textwidth]{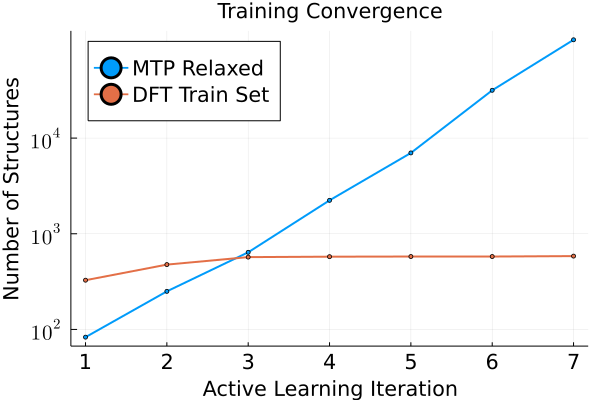}}
  \hfill
  \subfloat[]{\includegraphics[width=0.45\textwidth]{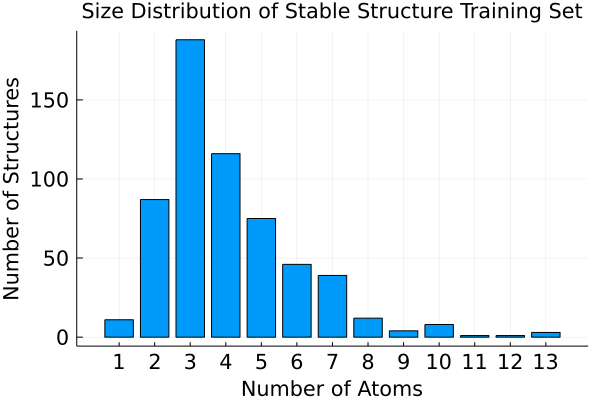}}
  \caption{(a) Early in training, the number of structures needed for the training set is significantly higher than the number of structures the MLIP is predicting on. As training progresses to larger supercells, however, the need for additional structures becomes negligible. (b) Thus, the size of the structures in the training set are dominated by small-cell structures.}
    \label{fig:convexSizes}
\end{figure}

Stability is assessed by comparing the formation enthalpy (at $T=0$ K) of a candidate structure with the formation enthalpy of structures that are known experimentally. Structures that are observed in experiment will lie on or near the lower convex hull of formation enthalpy formed by known stable structures \cite{nyshadham_computational_2017, levy_hafnium_2010, levy_uncovering_2010}. To increase the chances of finding new stable structures, the energy of candidate structures is minimized with respect to atom position and cell shape, a process called relaxation \cite{gubaev_accelerating_2019}. 

An efficient way to generate candidates for stability is to consider a few structures that are known to be stable and vary the hydrogen concentration ranges by creating vacancies. To create a set of candidate structures for training, we enumerate configurations\cite{morgan_generating_2017} of the $\delta$, $\gamma$, and $\epsilon$ phases while varying hydrogen concentration. Systematically enumerating possible structures is straightforward, but the number of these structures grows exponentially with cell size. Despite this large increase in the amount of information the potential must successfully model, small-cell training leads to a training set that efficiently represents these candidate structures and their relaxation trajectories.

\begin{figure}
    \centering
    \captionsetup{justification=raggedright}
    \includegraphics[width=\linewidth]{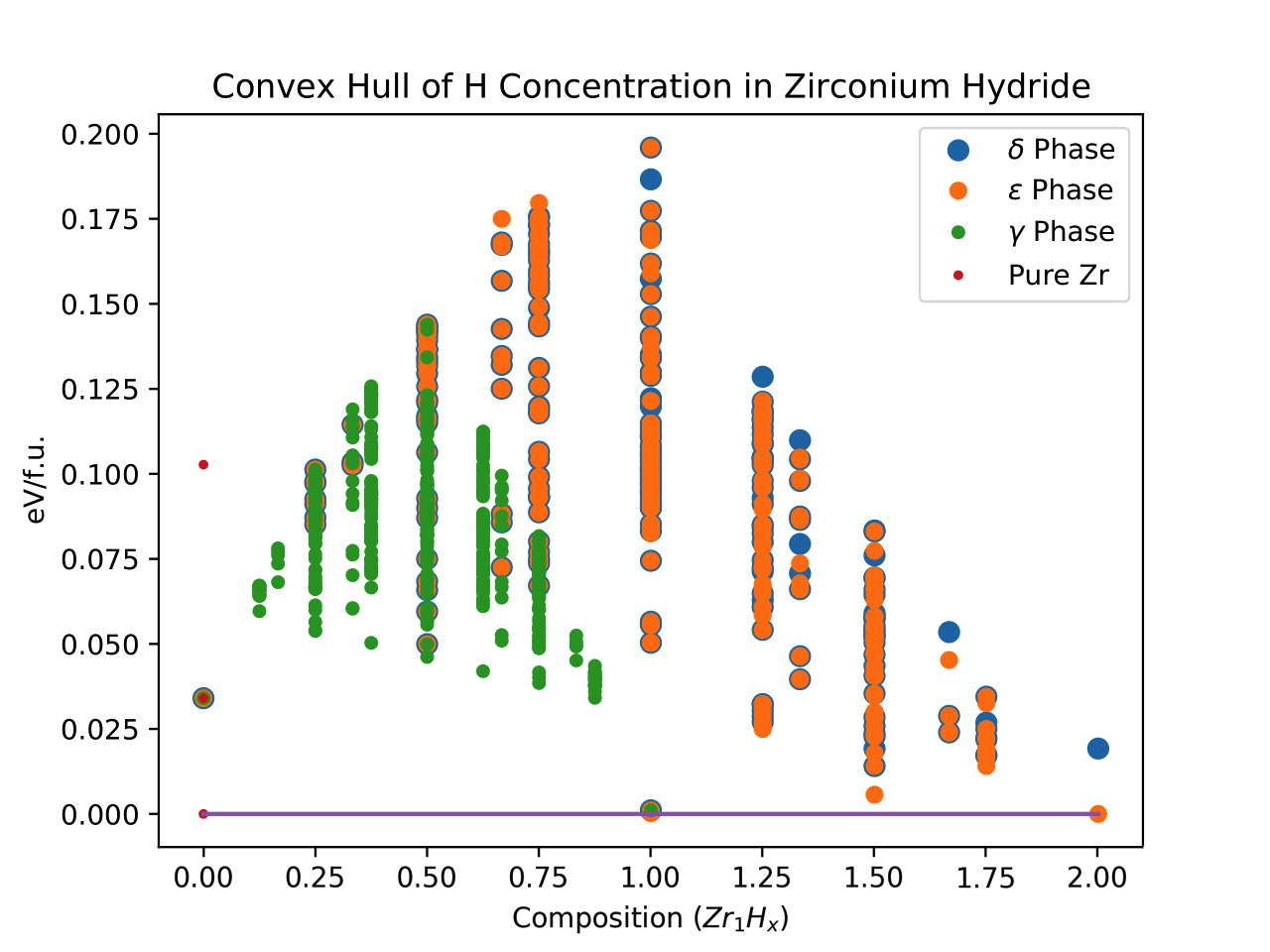}
    \caption{Small-cell training accurately reproduces the known convex hull of the Zr-H system. The $\alpha$ and $\epsilon$ phases are stable with $\gamma$ phase lying just above the convex hull. Small-cell training also effectively explores candidate structures for stability, finding low enthalpy structures that approach the convex hull at Zr\textsubscript{1}H\textsubscript{1.5}.}
    \label{fig:convex}
\end{figure}

Our iterative small-cell training began by applying active learning on both the relaxation trajectories of the primitive structures of the $\delta$, $\gamma$, and $\epsilon$ phases with varying hydrogen concentration and the relaxation trajectories of structures needed to compute the elastic constants of the $\delta$, $\gamma$, $\epsilon$, and $\alpha$ structure. The elastic constant structures were added to the initial set to create a more practical potential and to give the potential information about the $\alpha$ phase. As shown in Fig.~\ref{fig:convexSizes}(a), training with active learning required around 300 structures to be added to the training set before the potential could successfully relax the set of primitive structures and elastic constant structures. Once the MLIP could relax these structures without the active learning protocol requesting additional DFT calculations, training began on structures with the next smallest supercells. DFT calculations were done in VASP \cite{kresse_efficiency_1996, kresse_efficient_1996} using the parameters shown in Table \ref{tab:vasp}. The process was repeated for successively larger cells until the MLIP could successfully relax all of the enumerated structures. After finishing training, the potential could successfully relax 75032 structures while only having 584 structures in the training set (Fig.~\ref{fig:convexSizes}(a)). Additionally, 95\% of the structures in the training set had less than seven atoms.

\begin{figure}
  \centering
  \captionsetup{justification=raggedright}
  \subfloat[]{\includegraphics[width=0.5\textwidth]{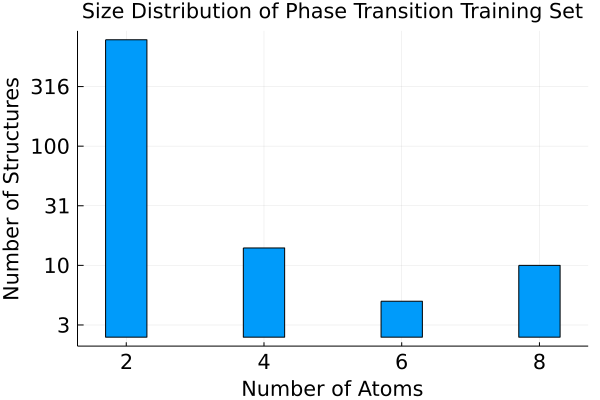}}
  \hfill
  \subfloat[]{\includegraphics[width=0.5\textwidth]{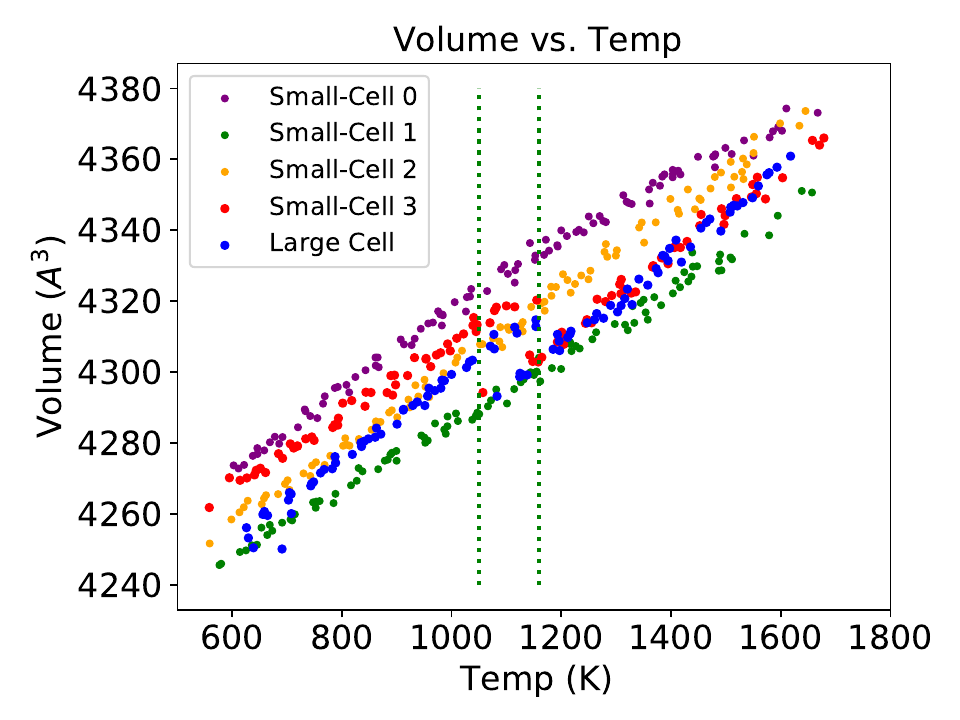}}
  \caption{(a) The size distribution of the training set for phase transition training is dominated by two atom structures. While structures of up to eight atoms are used, these are a small subset of the total training set. (b) As Small-Cell 0 shows, the small-cell trained potential initially fails to reproduce the phase transition or the thermal expansion of the $\alpha$ or $\beta$ phase (at temperatures above the correct phase transition temperature) that the large-cell trained potential (Large-Cell) successfully models. However, subsequent iterations of small-cell training first model the volume and thermal expansion correctly (Small-Cell 1 and Small-Cell 2) and finally are able to model the phase transition (Small-Cell 3). The vertical lines show that the phase transition temperature is in the same range for both the Small-Cell 3 potential and the Large-Cell potential.}
    \label{fig:transition}
\end{figure}

As training continued with larger supercells, the number of required DFT calculations decreased rapidly. After three increases in supercell size, the number of DFT calculations needed at each subsequent increase became nearly zero. The size distribution of the training set (Fig.~\ref{fig:convexSizes}(b)) shows that by the fourth increase, 95\% of the necessary DFT data had already been collected. Even though only a small number of structures are added to the training set during training with large supercells, the number of candidate structures that the potential successfully predicts continues to increase exponentially. This shows that prioritizing small-cell structures allows for most of the necessary information to be gathered in cheap, small-cell DFT calculations that give relevant information about the larger-cell structures, significantly decreasing the total training time. 

Additionally, using small-cell training still effectively represented the known convex hull well. The structures that sit on the convex hull are the $\alpha$ and $\epsilon$ phase, which matches DFT benchmarks. The $\gamma$ phase is also well represented, lying just above the convex hull. Small-cell training also successfully found candidates for stable structures at higher temperatures. Fig. ~\ref{fig:convex} shows structures lying close to the convex hull at Zr\textsubscript{1}H\textsubscript{1.5}. 

\begin{table}
\scriptsize{}
\captionsetup{justification=raggedright}
\centering
\begin{tabular}{c | c | c | c | c | c | c | c | c | c}
 & $C_{11}$ & $C_{22}$ & $C_{33}$ & $C_{44}$ & $C_{55}$ & $C_{66}$ & $C_{12}$ & $C_{13}$ & $C_{23}$ \\\hline
$\alpha$ (VASP) & 147.6 & 146.9 & 163.3 & 23.5 & 23.5 & 29.4 & 62.8 & 65.8 & 66.2 \\
$\alpha$ (MTP) & 142.0 & 142.0 & 165.3 & 28.7 & 28.7 & 28.6 & 84.8 & 66.4 & 66.4 \\\hline
$\delta$ (VASP) & 73.2 & 74.0 & 75.0 & 6.67 & 16.3 & 7.72 & 168.1 & 166.3 & 166.9 \\
$\delta$ (MTP) & 116.1 & 116.1 & 127.6 & -11.7 & -11.7 & -5.97 & 116.5 & 105.0 & 105.0 \\\hline
$\epsilon$ (VASP) & 190.2 & 186.0 & 168.1 & 29.0 & 41.0 & 29.4 & 104.3 & 120.6 & 110.7 \\
$\epsilon$ (MTP) & 207.1 & 197.7 & 266.8 & 39.9 & 41.4 & 40.4 & 133.4 & 132.4 & 131.3 \\\hline
$\gamma$ (VASP) & 193.0 & 193.0 & 179.1 & 44.8 & 44.8 & 0.43 & 61.8 & 88.2 & 88.2 \\
$\gamma$ (MTP) & 167.8 & 167.8 & 187.8 & 48.0 & 48.0 & -2.89 & 57.8 & 105.4 & 105.4
\end{tabular}
\caption{\label{tab:elast_initial}Elastic constants for $\alpha$, $\delta$, $\epsilon$, and $\gamma$ phases. Elements of the elastic tensor are in units of GPa.}
\begin{tabular}{c | c | c | c | c | c | c | c | c}
  & $\text{G}_\text{V}$ & $\text{G}_\text{R}$ & $\text{G}_\text{VRH}$ & $\text{K}_\text{V}$ & $\text{K}_\text{R}$ & $\text{K}_\text{VRH}$ & $\nu$ (unitless) & $E$ \\\hline
 $\alpha$ (VASP) & 32.8 & 30.3 & 31.6 & 94.5 & 94.2 & 94.3 & 0.35 & 85.2 \\
 $\alpha$ (MTP) & 32.7 & 31.2 & 31.9 & 98.3 & 98.3 & 98.3 & 0.35 & 86.4 \\\hline
 $\delta$ (VASP) & -12.5 & 2.24 & -5.12 & 135.7 & 135.7 & 135.7 & 0.52 & -15.5 \\
 $\delta$ (MTP) & -3.67 & 24.9 & 10.6 & 112.5 & 112.5 & 112.5 & 0.45 & 30.9 \\\hline
 $\epsilon$ (VASP) & 33.8 & 30.0 & 31.9 & 136.1 & 129.0 & 132.5 & 0.39 & 88.6 \\
 $\epsilon$ (MTP) & 42.6 & 37.6 & 40.1 & 162.9 & 151.4 & 157.1 & 0.38 & 110.9 \\\hline
 $\gamma$ (VASP) & 39.8 & 2.07 & 20.9 & 118.1 & 118.0 & 118.1 & 0.42 & 59.3 \\
 $\gamma$ (MTP) & 35.6 & -19.9 & 7.83 & 117.9 & 112.2 & 115.0 & 0.47 & 23.0 \\\hline
\end{tabular}
\caption{\label{tab:relax_prop_elast}Elastic properties for $\alpha$, $\delta$, $\epsilon$, and $\gamma$ phases. Unless otherwise noted, reported elastic properties are in units of GPa. $G$ and $K$ represent bulk modulus and shear modulus respectively, while $V$, $R$, and $H$ represent Voigt, Reuss, and Hill respectively. $\nu$ is Poisson's ratio and $E$ is Young's modulus.}
\end{table}

\section{Small-cell Training For Large-scale Simulations}

\begin{figure*}
  \centering
  \captionsetup{justification=raggedright}
  \subfloat[$\alpha$ phase]{\includegraphics[width=0.5\linewidth]{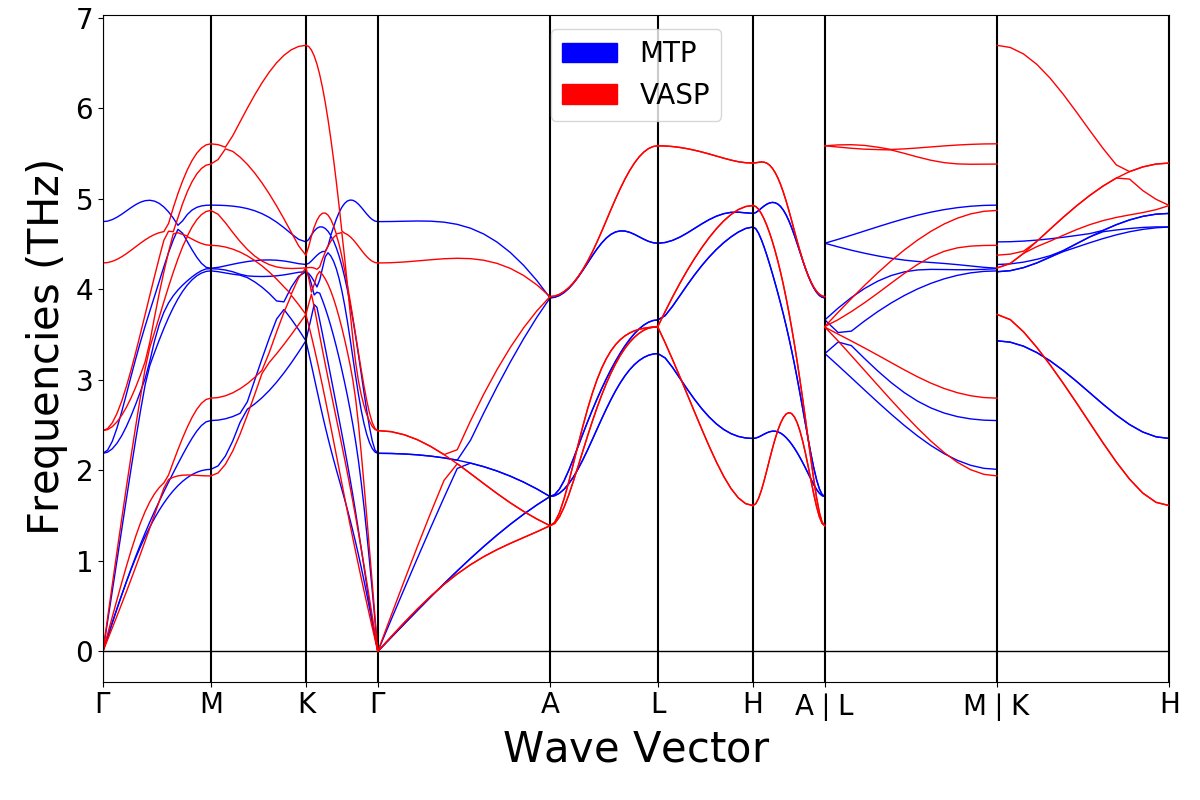}}
  \hfill
  \subfloat[$\delta$ phase]{\includegraphics[width=0.5\linewidth]{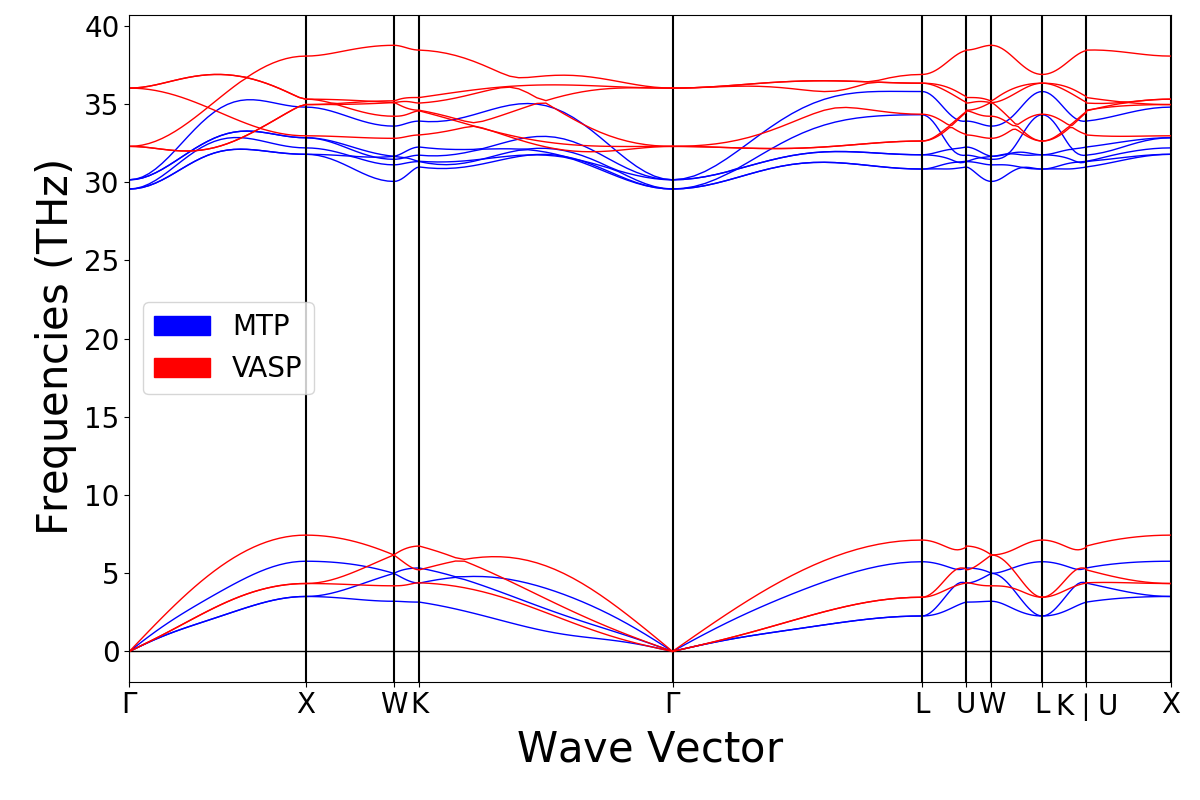}}
  \hfill
  \subfloat[$\epsilon$ phase]{\includegraphics[width=0.5\linewidth]{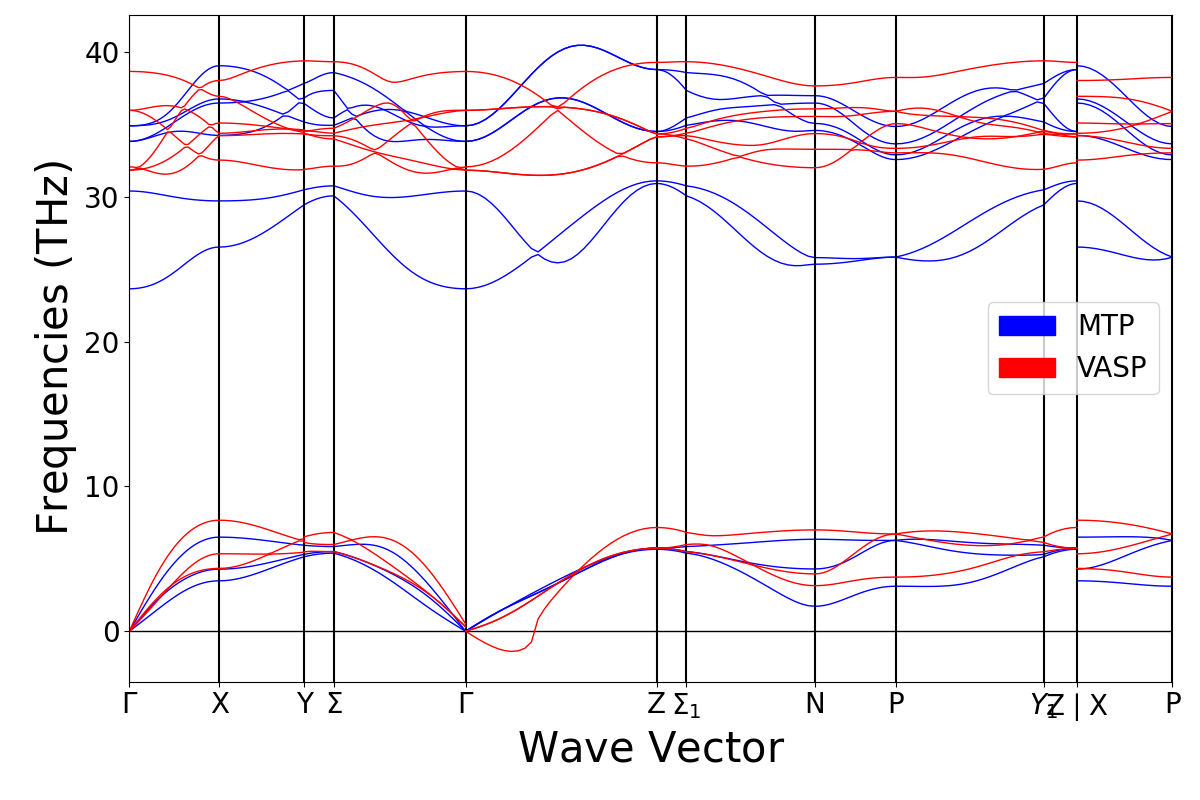}}
  \hfill
  \subfloat[$\gamma$ phaes]{\includegraphics[width=0.5\linewidth]{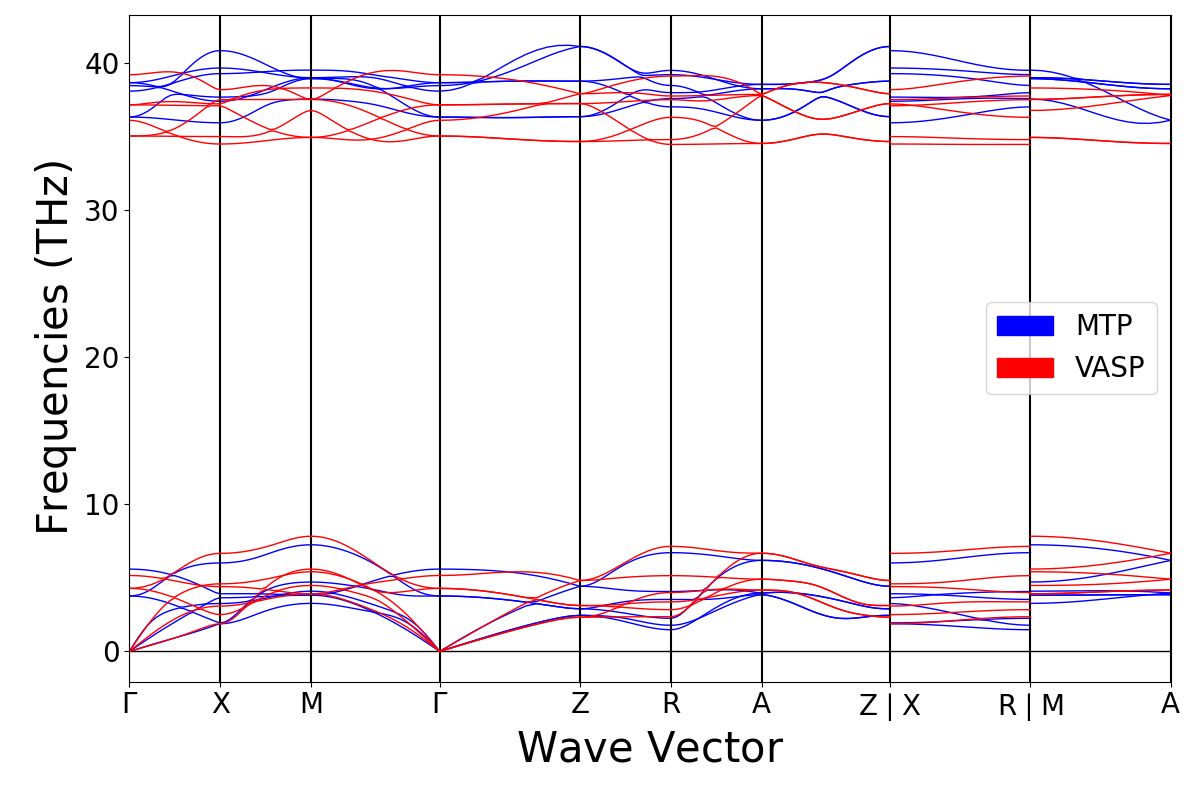}}
\caption{\label{fig:phonon}Phonon band structures for $\alpha$, $\delta$, $\epsilon$, and $\gamma$ phases.}
\end{figure*}

When searching for stable structures of the Zr-H system, candidate structures for stability could be ordered by size to ensure that training starts with the small-cell structures, making small-cell training suitable. In contrast, a large-scale simulation can require hundreds or thousands of atoms for the physical phenomena to be apparent, making small-cell training seem inappropriate. However, small-cell training can be used to create potentials that model large-scale simulations if we train the potential with a sequence of simulations that increase in cell size and converge to the full simulation.

For example, a phase transition simulation is a large-scale simulation that often requires simulation cells with several hundred atoms. However, instead of starting training with a large-cell structure with hundreds of atoms, the primitive structure is used. When the potential can successfully finish this small-cell phase transition simulation without further DFT requests, the supercell size is increased to twice the primitive. Training progresses with larger and larger supercells until the potential can successfully capture the phase transition in a large-scale simulation.

\begin{figure}
  \centering
  \captionsetup{justification=raggedright}
  \subfloat[DFT Prediction]{\includegraphics[width=\linewidth]{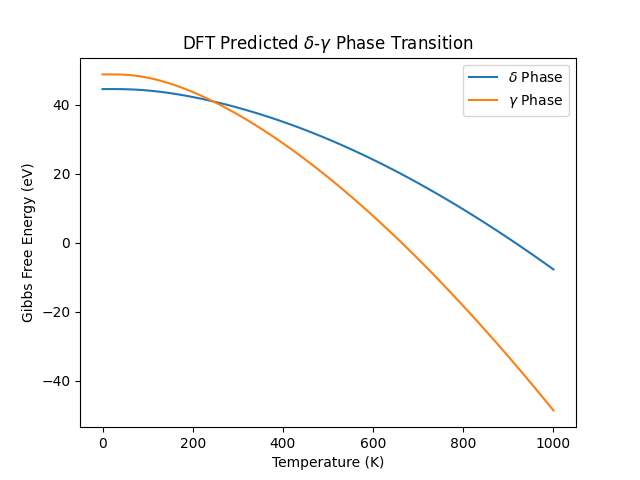}}
  \hfill
  \subfloat[MTP Prediction]{\includegraphics[width=\linewidth]{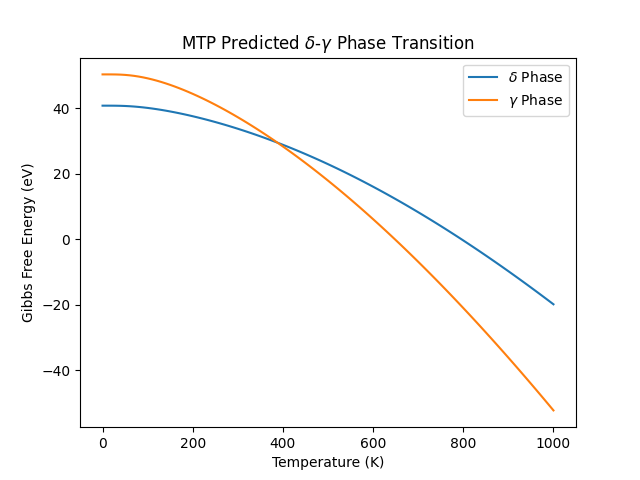}}
\caption{\label{fig:qha}Predictions of the $\delta$-$\gamma$ phase transition temperature using the quasi-harmonic approximation for Gibbs free energy.}
\end{figure}

From a physical perspective, this method is counter-intuitive. A phase transition simulation of only a few atoms cannot capture the phase transition and does not effectively model experimental conditions. From the perspective of taking a limit, however, the idea is reasonable. As training progresses with larger and larger supercells, the simulations used for training will more accurately approximate the large-scale simulation. Still, it is unclear how effective small-cell training is when some of the simulations used in training do not accurately model reality.

To assess the effectiveness of small-cell training for large-scale simulations, we consider the zirconium $\alpha$-$\beta$ phase transition. The $\alpha$-$\beta$ phase transformation of pure zirconium is well characterised and known experimentally, however the transformation is strongly affected by alloying elements \cite{Yi2006}. Understanding the impact of alloying on this phase transformation is important both in terms of optimising alloying processing and to understand component integrity during reactor loss-of-coolant-accidents (LOCA); future work could take the approach described here and apply it to determine the impact of alloying elements on phase transformation.

Because both the $\alpha$-$\beta$ phase transition and the Zr-H system are important for modeling zirconium in nuclear environments, the earlier potential developed to explore stable structures of the Zr-H system was further trained for modeling the $\alpha$-$\beta$ phase transition. In the first iteration of small-cell training, a two-atom heating simulation of $\alpha$ phase and a two-atom cooling simulation of $\beta$ phase were performed in LAMMPS \cite{thompson_lammps_2022} with active learning. After these simulations ran successfully without requiring additional DFT calculations, small-cell training moved to four-atom simulations for both the heating and cooling simulations. The size of the supercells increased until the potential could model the full-scale phase transition simulation, finding that the potential captured the phase transition after training on molecular dynamics temperature ramps containing only eight atoms (Fig.~\ref{fig:transition}(a)). Fig.~\ref{fig:transition}(b) shows the effect of small-cell training on the potential. The potential initially fails to reproduce the phase transition and the thermal expansion of the $\alpha$ and $\beta$ phases for temperatures above the phase transition temperaaturees (Small-Cell 0). However, as further iterations of small-cell training are completed, the thermal expansion of $\alpha$ (Small-Cell 1) and $\beta$ (Small-Cell 2) are successfully modeled. After training is finished, the potential is able to model both the phase transition and the thermal expansion of the $\alpha$ and $\beta$ phases (Small-Cell 3). DFT calculations were performed in VASP with the same parameters used earlier, as defined in Table \ref{tab:vasp}.

It may be surprising that the potential can model the phase transition despite the training set containing only small structures. A simulation with only eight atoms cannot model the $\alpha$-$\beta$ phase transition, even with a perfectly accurate interatomic potential, due to finite-size effects. However, these small-cell simulations provide a large and diverse sample of atomic environments similar to those that would be encountered by the large-scale phase transition simulation. After training, the local environments in the training set are diverse enough that the large-scale phase transition simulation is unlikely to encounter an atomic environment that is not well represented in the training set.

To evaluate the accuracy of the phase transition, we compare it to a second MTP that accurately modeled the phase transition but trained with large supercells \cite{peitao2021}. The initial state of the training set of the second potential included large-cells with random atomic displacements. Then, the second potential was trained directly on the full-scale simulation with an active learning strategy. After three iterations of training, the MTP successfully modeled the phase transition between $\alpha$ and $\beta$ phases of zirconium. Convergence was achieved within four iterations, giving rise to a training set containing a total number of 653 structures, which consist of 330 structures of $\alpha$ phase (54 atoms per structure) and 323 structures of $\beta$ phase (48 atoms per structure), respectively.\footnote{The parameters used for the VASP calculations of this training set were not the same as the parameters shown in Table~\ref{tab:vasp}. However, the only significant parameter change from the parameters in Table~\ref{tab:vasp} was a change in ENCUT from 400 eV to 500 eV. Because 400 eV is a reasonable value \fixes{}{for }of ENCUT for the Zr system, this should not affect the training results significantly.}

When compared with the MTP trained with large supercells, the phase transition of our MTP has good agreement. The phase transition temperature is in the same range for both and the thermal expansion coefficient for the $\beta$ phase also agrees. The thermal expansion coefficient for $\alpha$ phase is slightly lower with our MLIP, but for a given temperature, the volume is accurate to within 1\% (Fig. ~\ref{fig:transition}). In addition to retaining the accuracy of the MLIP trained on large-cells, our small-cell trained potential also had a large reduction in training time. 

The total cpu time used for DFT calculations during the training of our potential was 11.5 times less than used for the large-cell training. However, this speedup includes calculations from training for both the phase transition and when searching for stable Zr-H structures. When restricted to cpu time spent for training on the phase transition, the cpu time for DFT calculations is 17.5 times less than large-cell training.

\section{Z\lowercase{r}-H Potential Analysis}

The Zr-H potential developed in this paper was trained to model a diverse set of candidates for thermodynamic stability, a few elemental structures, and the pure zirconium $\alpha$-$\beta$ phase transition. However, we wish to evaluate the potentials ability to model several other material properties such as the relaxed state, phonon bandstructure, and elastic properties of the $\alpha$, $\delta$, $\epsilon$, and $\gamma$ phases. All DFT predictions were calculated with the VASP parameters shown in Table~\ref{tab:vasp}.

The predictions of the relaxed state of the $\alpha$, $\delta$, $\epsilon$, and $\gamma$ phases is depicted in Table~\ref{tab:relax_prop_initial}. These predictions are highly accurate, with less than a 1\% relative error for all predictions. Given that the relaxation trajectories were included in the training when searching for thermodynamically stable structures, this may not be surprising. However, it does increase the confidence that the predictions of the convex hull are correct for these phases.

Using pymatgen\cite{ong_python_2013}, predictions of the elastic properties of the four phases were also reproduced with reasonable accuracy. Tables~\ref{tab:elast_initial} and~\ref{tab:relax_prop_elast} shows that the elastic constants and other elastic properties of the $\alpha$ and $\gamma$ phases are reproduced well. However, the potential fails to reproduce elastic properties for the $\delta$ and $\epsilon$ phases to high accuracy, often exceeding 30 GPa in absolute error.

As Fig.~\ref{fig:phonon} shows, the potential is able to \fixes{reproduce}{predict} the general trend of the \fixes{DFT predicted }{}phonon bandstructures. However, for the \fixes{$\epsilon$}{$\delta$} phase, the acoustic branch \fixes{fails to recreate}{predicts} an imaginary mode instability \fixes{that DFT predicts}{where it should be stable}. Additionally, the optical branches of the phonon bandstructures are underestimated by the MTP.

Ignoring configurational entropy, the potential also predicts the existence of the Zr\textsubscript{1}H\textsubscript{2} $\delta$ to Zr\textsubscript{2}H\textsubscript{2} $\gamma$ phase transition using the quasi-harmonic approximation for Gibbs free energy \fixes{ (Fig.~\ref{fig:qha})}{}. However, the predicted MTP phase transition temperature is inconsistent with DFT, estimating a phase transition temperature of $\sim$150 K higher. Both the phonon bandstructures and estimated phase transition temperature were calculated with phonopy \cite{togo_implementation_2023}.

\section{Conclusion}

While active learning presents a methodical way to create a training set for complex material properties, it is often inefficient. Our small-cell training protocol allows inexpensive, small-cell DFT calculations to give information about large-cell structures with reasonable accuracy, avoiding computationally expensive calculations with larger supercells. 

Small-cell training has the potential to improve other active learning approaches as well. Other active learning methods can incorporate small-cell training into their frameworks. Small-cell training only changes large simulations to small simulations when training, making it indifferent to the type of active learning used.

Small-cell training can effectively create a robust and comprehensive potential for modeling complex material properties in the Zr-H system. Our potential can model the relaxed state, elastic constants, and phonon bandstructures of the $\alpha$ phase and the hydride $\epsilon$, $\delta$, and $\gamma$ phases as well as several candidates for stable structures at higher temperatures. Additionally, this potential can accurately model the $\alpha$-$\beta$ phase transition, reproducing the thermal expansion coefficients and the phase transition temperature. Small-cell training also models these material properties more efficiently than typical active learning methods alone. When compared with a potential trained on large supercells, training took up to 17.5 times longer.

\section{Declarations}

\subsection{Funding}

This work was funded by the Advanced Materials Simulation Engineering Tool (AMSET) project, sponsored by the US Naval Nuclear Laboratory (NNL) and directed by Materials Design, Inc. YL, MRD, and LKB thank the Digital Research Alliance of Canada (formerly known as Compute Canada) for generous allocation of compute resources, and the Natural Sciences and Engineering Research Council of Canada (NSERC) and the NSERC/UNENE Industrial Research Chair in Nuclear Materials at Queen’s for financial support.

\subsection{Conflicts of interest/Competing interests}

On behalf of all authors, the corresponding author states that there is no conflict of interest.

\subsection{Availability of data and material}

The moment tensor potential file (pot.mtp) and the training set file (train.cfg) are available at https://github.com/jmeziere/Zr-H-Potential.

\bibliography{main}

\section{Authors Contributions}

\textbf{Jason Meziere}: Conceptualization, Methodology, Formal analysis and investigation, Writing - original draft preparation, Writing - review and editing. 
\textbf{Yu Luo}: Formal analysis and investigation, Writing - review and editing. 
\textbf{Yi Xia}: Formal analysis and investigation, Writing - review and editing. 
\textbf{Laurent Karim Beland}: Formal analysis and investigation, Writing - review and editing.
\textbf{Mark Daymond}: Formal Analysis and investigation, Writing - review and edition, Funding acquisition.
\textbf{G L W Hart}: Conceputalization, Methodology, Formal analysis and investigation, Writing - review and edition, Funding acquisition, Resources, Supervision

\onecolumngrid

\end{document}